\documentclass[]{elsarticle}

\usepackage{hyperref}
\usepackage{subfigure}

\journal{Journal}

\graphicspath{{figs/}}

\begin{document}

\begin{frontmatter}

\title{Living mycelium composites discern weights}

%% Authors and affiliations:
\author[1]{Andrew Adamatzky}
\author[2]{Antoni Gandia}

\address[1]{Unconventional Computing Laboratory, UWE, Bristol, UK}
\address[2]{Institute for Plant Molecular and Cell Biology, CSIC-UPV, Valencia, ES}

\begin{abstract}
Fungal construction materials --- substrates colonised by mycelium --- are getting increased recognition as viable ecologically friendly alternatives to conventional building materials. A functionality of the constructions made from fungal materials would be enriched if blocks with living mycelium, known for their ability to respond to chemical,  optical and tactile stimuli, were inserted. We investigate how large blocks of substrates colonised with mycelium of \emph{Ganoderma resinaceum} respond to stimulation with heavy weights. We analyse details of the electrical responses to the stimulation with weights and show that ON and OFF stimuli can be discriminated by the living mycelium composites and that a habituation to the stimulation occurs. 
\end{abstract}


\end{frontmatter}

%\linenumbers

\section{Introduction}

Mycelium bound composites --- masses of organic substrates colonised by fungi --- are considered to be future environmentally sustainable growing biomaterials~\cite{karana2018material,jones2020engineered,cerimi2019fungi}. The fungal materials are used in acoustic insulation panels~\cite{pelletier2013evaluation,elsacker2020comprehensive,robertson2020fungal}, thermal insulation wall cladding~\cite{yang2017physical,xing2018growing,girometta2019physico,dias2021investigation,wang2016experimental,cardenas2020thermal}, packaging materials~\cite{holt2012fungal,sivaprasad2021development,mojumdar2021mushroom} and wearables~\cite{adamatzky2021reactive,silverman2020development,karana2018material,appels2020use,jones2020leather}. 

In \cite{adamatzky2019fungal} we proposed to develop a structural substrate by using live fungal mycelium, functionalise the substrate with nanoparticles and polymers to make mycelium-based electronics~\cite{beasley2020capacitive,beasley2020mem,beasley2020fungal}, implement sensorial fusion and decision making in the mycelium networks~\cite{adamatzky2020boolean} and to grow monolithic buildings from the functionalised fungal substrate~\cite{adamatzky5adaptive}. 
Fungal buildings would self-grow, build, and repair themselves subject to substrate supplied, use natural adaptation to the environment, sense all that humans can sense. Whilst major parts of a building will be made from dried and cured mycelium composites there is an opportunity to use blocks with living mycelium as embedded sensorial elements. On our venture to investigate sensing properties of the mycelium composite blocks, called `fungal blocks' further, we decided to study how large fungal blocks respond to pressure via changes in their electrical activity. The electrical activity have been chosen as indicator because fungi are known to respond to chemical and physical stimuli by changing patterns of their electrical activity~\cite{olsson1995action,adamatzky2018spiking,adamatzky2018towards} and electrical properties~\cite{beasley2020fungal}. 

In Sect. 1 we describe experimental setup used to record electrical activity of mycelium bound composites and to stimulate the fungal blocks. Electrical responses of the fungal blocks to stimulation with weights are analysed in Sect. 2. Section 3 outlines our view on future works in the field.

\section{Methods}

\begin{figure}[!tbp]
    \centering
    \subfigure[]{\includegraphics[width=0.3\textwidth]{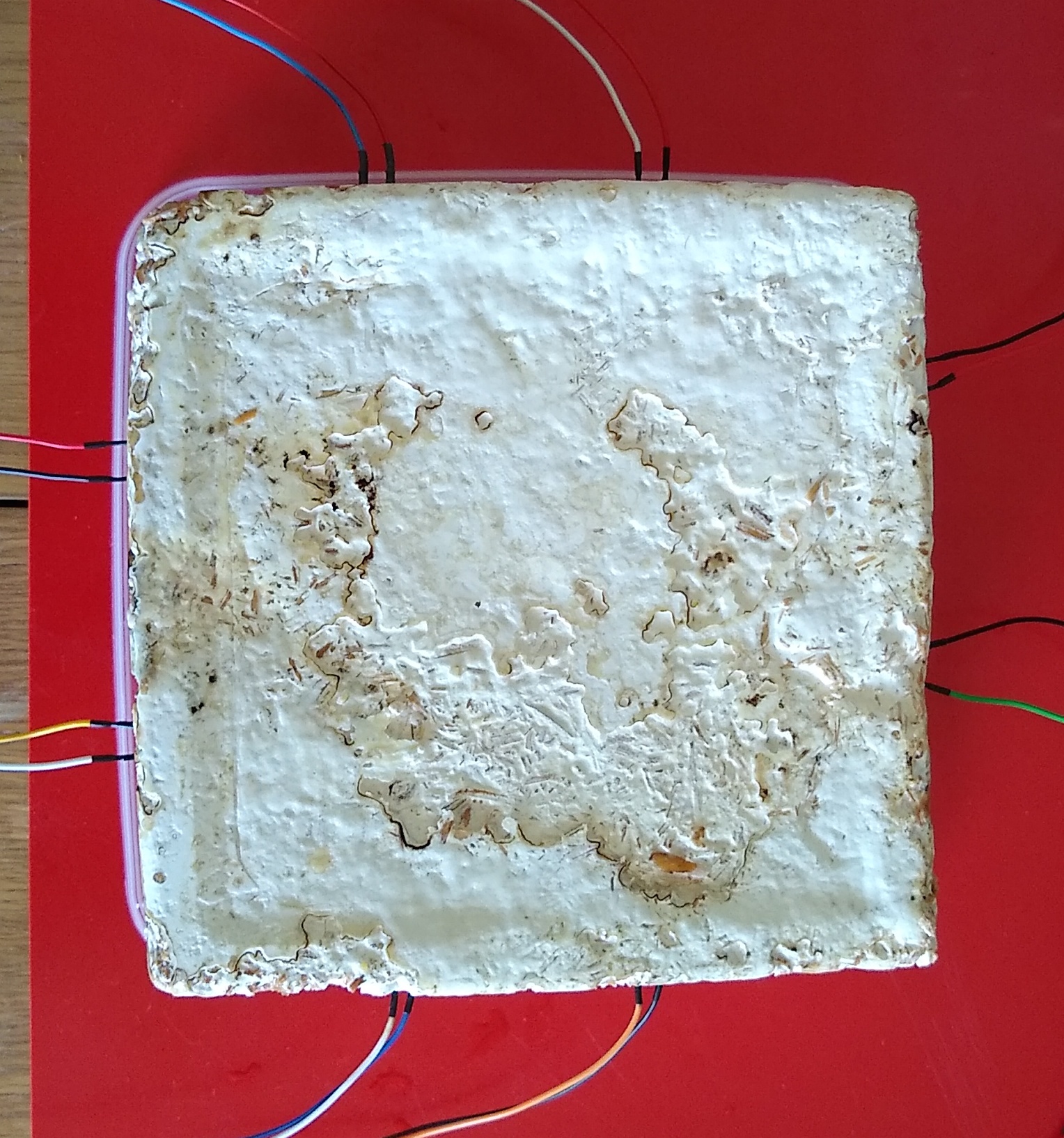}}
    \subfigure[]{\includegraphics[width=0.39\textwidth]{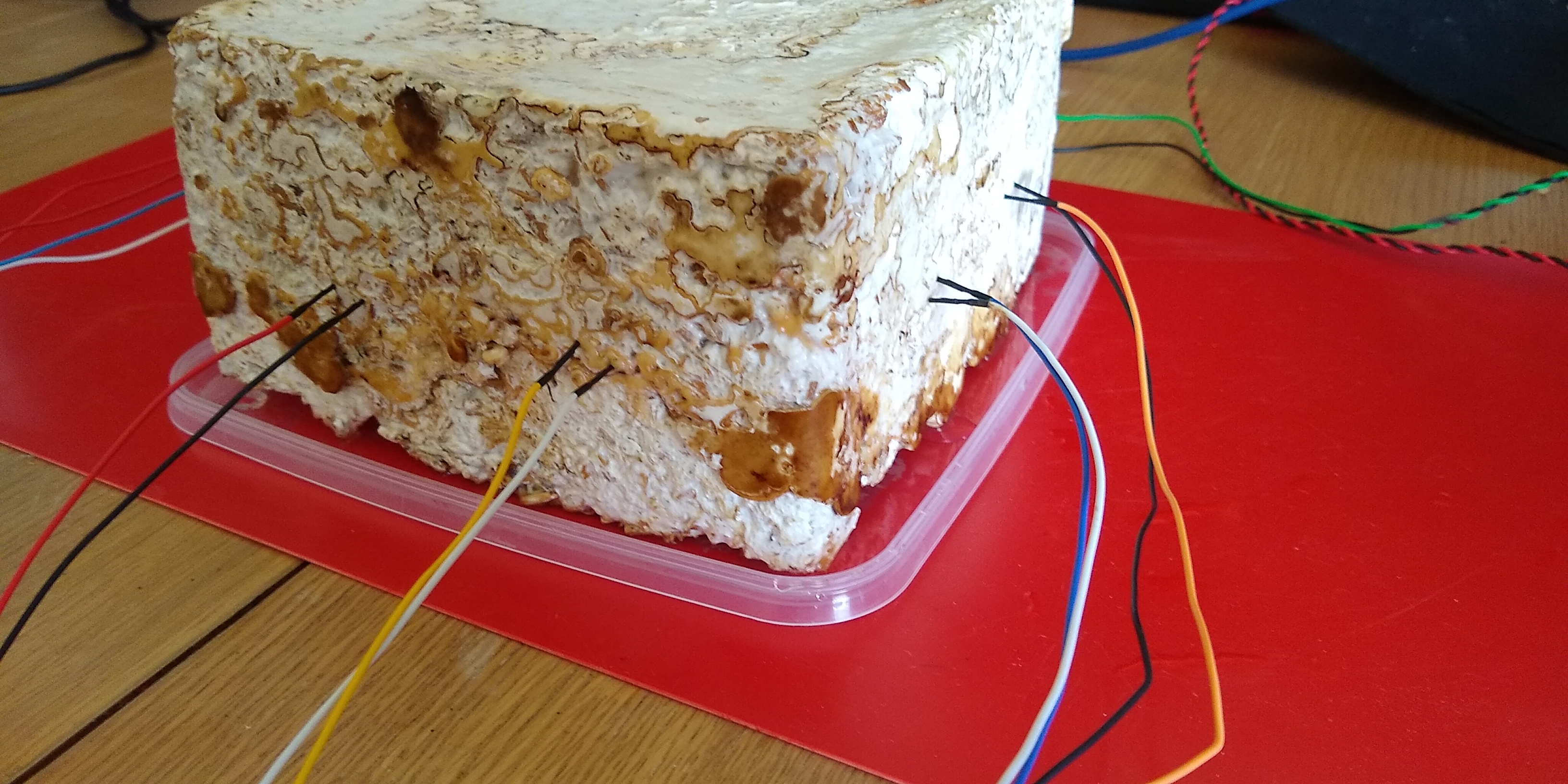}}
 \subfigure[]{\includegraphics[width=0.29\textwidth]{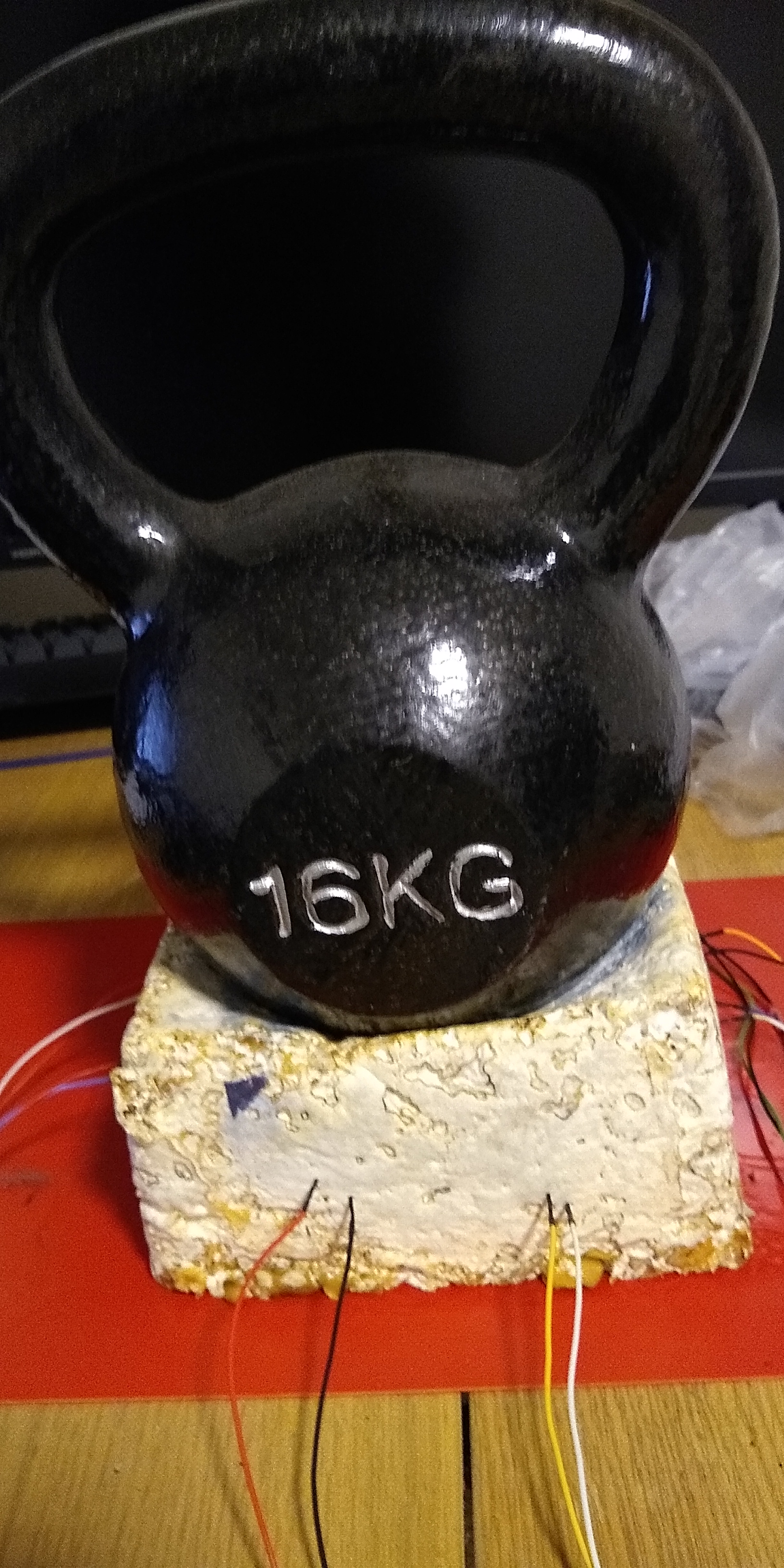}}
    \caption{Experimental setup. (ab)~Position of electrodes in fungal blocks: (a)~top view and (b)~side view. (c)~Pairs of differential electrodes inserted in a fungal block and 16~kg kettle bell placed on top of the fungal block.}
    \label{fig:setup}
\end{figure}

A strain of the filamentous polypore fungus \emph{Ganoderma resinaceum} (MOGU's collection code 19-18, Mogu S.r.l., Inarzo, Italy), pre-selected for its superior fitness growing on the targeted substrate, was cultured on a substrate based of hemp shives and soybean hulls in plastic filter-patch microboxes in darkness at ambient room temperature c.~22\textsuperscript{o}C. Living blocks of the colonised substrate c.~20$\times$20$\times$10~cm were used for the experiments.

Electrical activity of the colonised fungal blocks was recorded using pairs of iridium-coated stainless steel sub-dermal needle electrodes (Spes Medica S.r.l., Italy), with twisted cables and  ADC-24 (Pico Technology, UK) high-resolution data logger with a 24-bit A/D converter, galvanic isolation and software-selectable sample rates all contribute to a superior noise-free resolution.  The pairs of electrodes were pierced into sides of the blocks as shown in Fig.~\ref{fig:setup}ab, two pairs per side. Distance between electrodes was 1-2~cm. In each trial, we recorded 8 electrode pairs, channels, simultaneously. We recorded electrical activity one sample per second. During the recording, the logger has been doing as many measurements as possible (typically up to 600 per second) and saving the average value. The humidity of the fungal blocks was 70\%-80\% (MerlinLaser Protimeter, UK). The experiments were conducted in a growing tent with constant ambient temperature at 21\textsuperscript{o}C in absence of light.  

We stimulated the fungal blocks by placing a 8~kg and 16~kg cast iron weights on the tops of the blocks (Fig.~\ref{fig:setup})c. The surface of the fungal blocks was insulated from the cast iron weight by a polyethylene film.

\section{Results}

\begin{figure}[!tbp]
    \centering
    \subfigure[]{\includegraphics[width=0.9\textwidth]{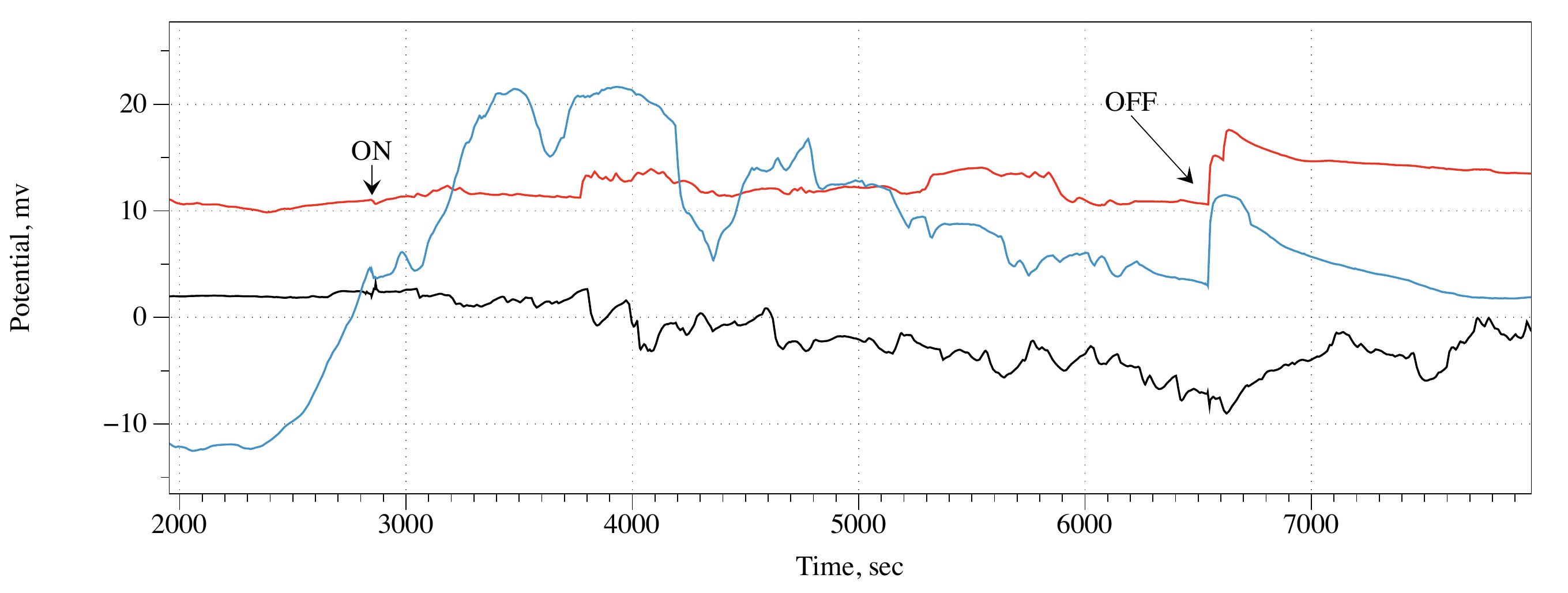}}
   \subfigure[]{\includegraphics[width=0.99\textwidth]{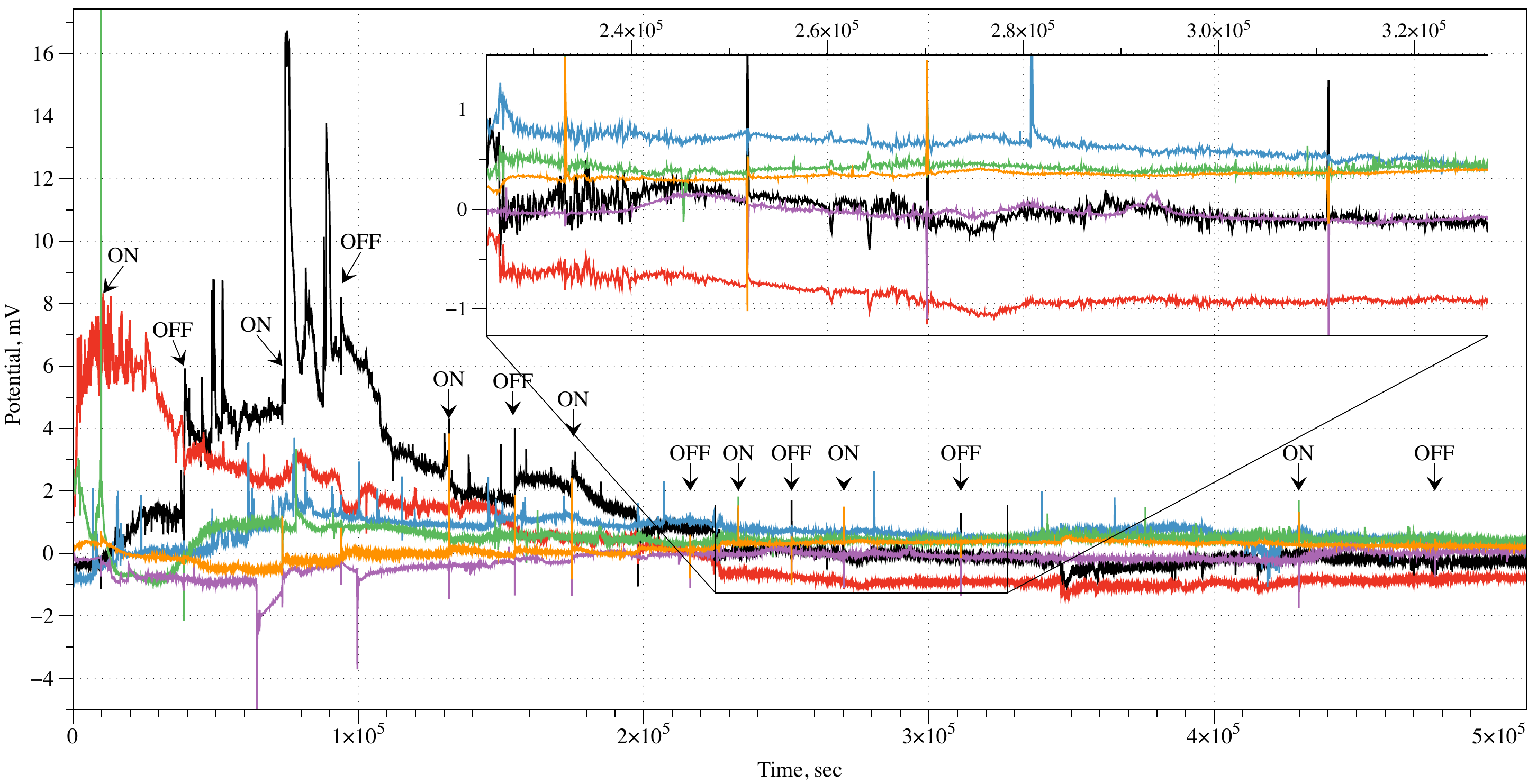}}
    \caption{Electrical activity of the fungal blocks, stimulated with heavy loads. (a)~The activity of the block stimulated with 8~kg load. (b)~The activity of the block stimulated with 16~kg load.  Moments of the loads applications are labelled by `ON' and lifting the loads by `OFF'.}
    \label{fig:examples}
\end{figure}

An example of fungal block's electrical responses to 8~kg load is shown in Fig.~\ref{fig:examples}a. The responses are characterised by an immediate response, i.e. occurring in 10-20~min of the stimulation, and a delayed, in 1-4~hours after beginning of the stimulation, response. The immediate responses were manifested in spikes of electrical potential recorded on the electrodes. Average amplitude of an immediate response to the loading with 8~kg was 3.05~mV, $\sigma=2.5$ and the spikes' average duration 489~sec, $\sigma=273$. Average amplitude of the immediate response to lifting the weight was 4~mV, $\sigma=4.4$ and average duration of 217~sec, $\sigma=232$. Delayed responses were manifested in trains of spikes with average amplitude 1.7~mV, $\sigma=1$, average duration of 161~sec, $\sigma=74$ and distance between spikes 125~sec, $\sigma=34$.

\begin{figure}[!tbp]
    \centering
    \subfigure[]{\includegraphics[width=0.8\textwidth]{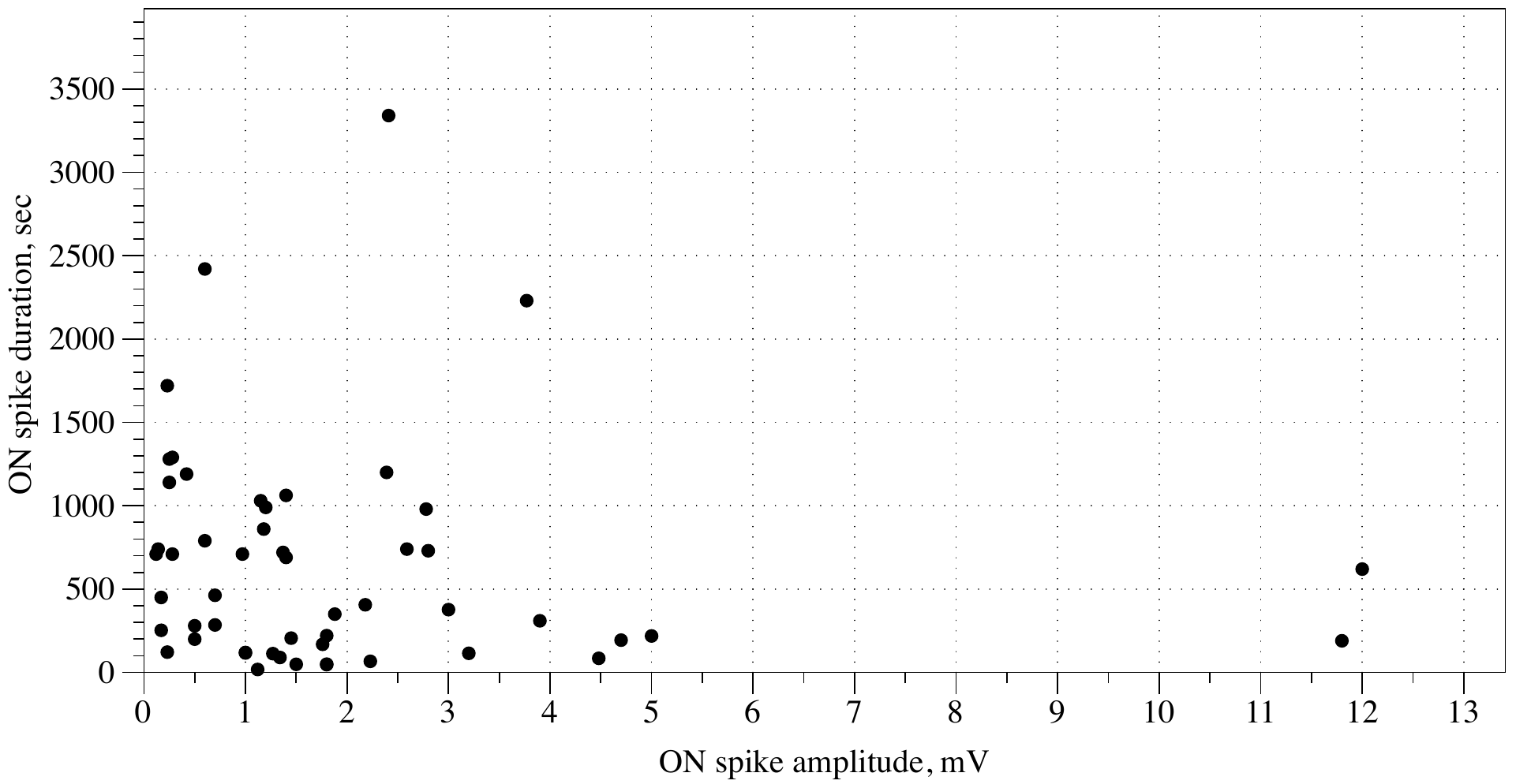}}
   \subfigure[]{\includegraphics[width=0.8\textwidth]{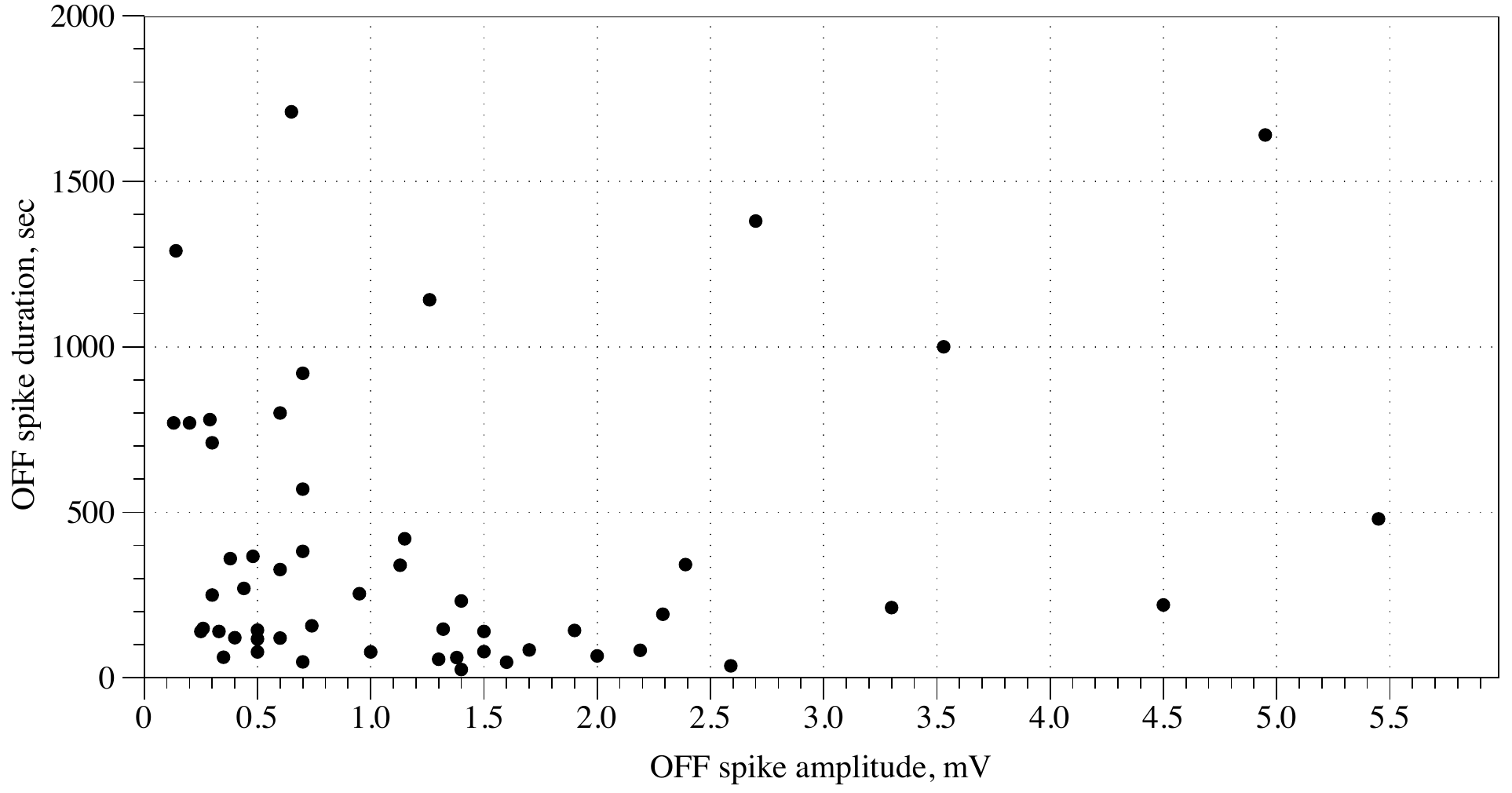}}
    \caption{Distribution of spike-response amplitude versus duration in response to (a)~application of the 16~kg weight and (b)~removal of the 16~kg weight.}
    \label{fig:distributions}
\end{figure}

Fungal blocks shown pronounced responses to the 8~kg loads only for 1-2 cycles of loading and unloading, no significant responses to further cycles of the stimulation have been observed.  The fungal blocks responded to stimulation with 16~kg weight for at least 8 cycles of loading and unloading. Let us discuss these responses in details. 

An example of electrical activity recorded on 8 channels, during the stimulation with 16~kg weight, is shown in Fig.~\ref{fig:examples}b. Distributions of `spikes amplitudes versus spikes durations' for spike-responses to application of the weight (ON spikes) and lifting of the weight (OFF spikes) are shown in Fig.~\ref{fig:distributions}. In response to application of 16~kg weight the fungal blocks produced spikes with median amplitude 1.4~mV and median duration 456~sec; average amplitude of ON spikes was 2.9~mV, $\sigma=4.9$ and average duration 880~sec, $\sigma=1379$. OFF spikes were characterised by median amplitude 1~mV and median duration 216~sec; average amplitude 2.1~mV, $\sigma=4.6$, and average duration 453~sec, $\sigma=559$. ON spikes are 1.4 higher than and twice as longer as OFF spikes. Based on this comparison of the response spikes we can claim that fungal blocks recognise when a weight was applied or removed. 

\begin{figure}[!tbp]
    \centering
    \subfigure[]{\includegraphics[width=0.49\textwidth]{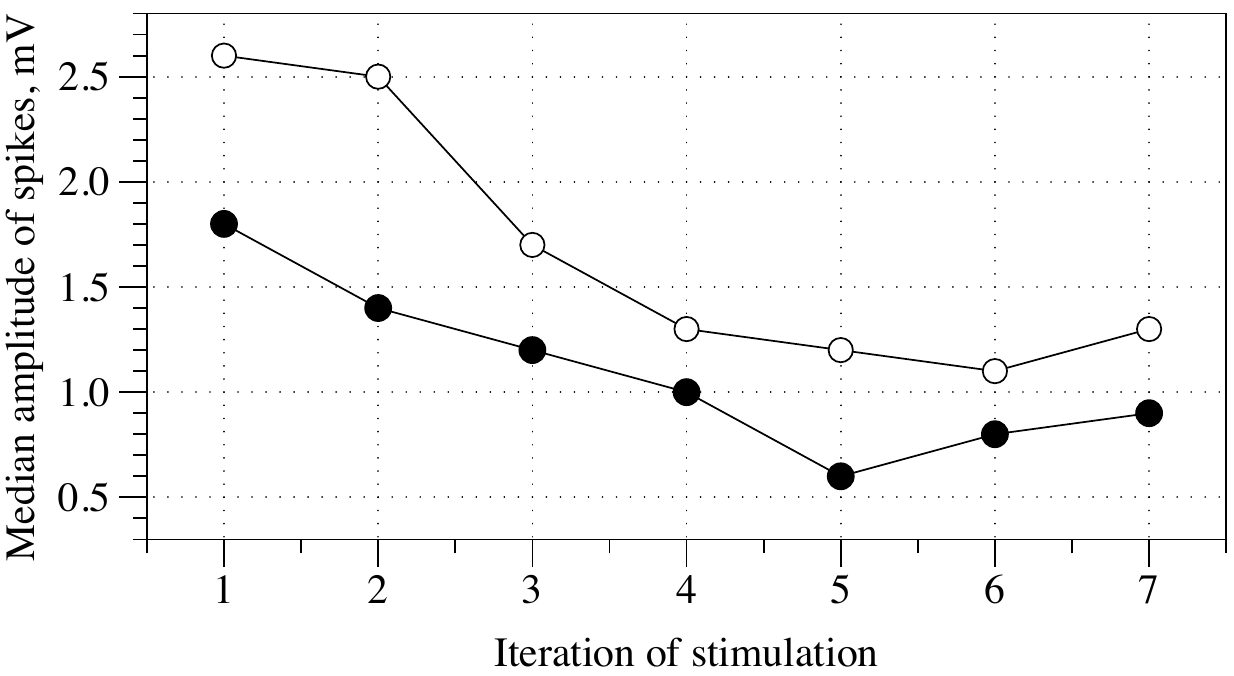}}
   \subfigure[]{\includegraphics[width=0.49\textwidth]{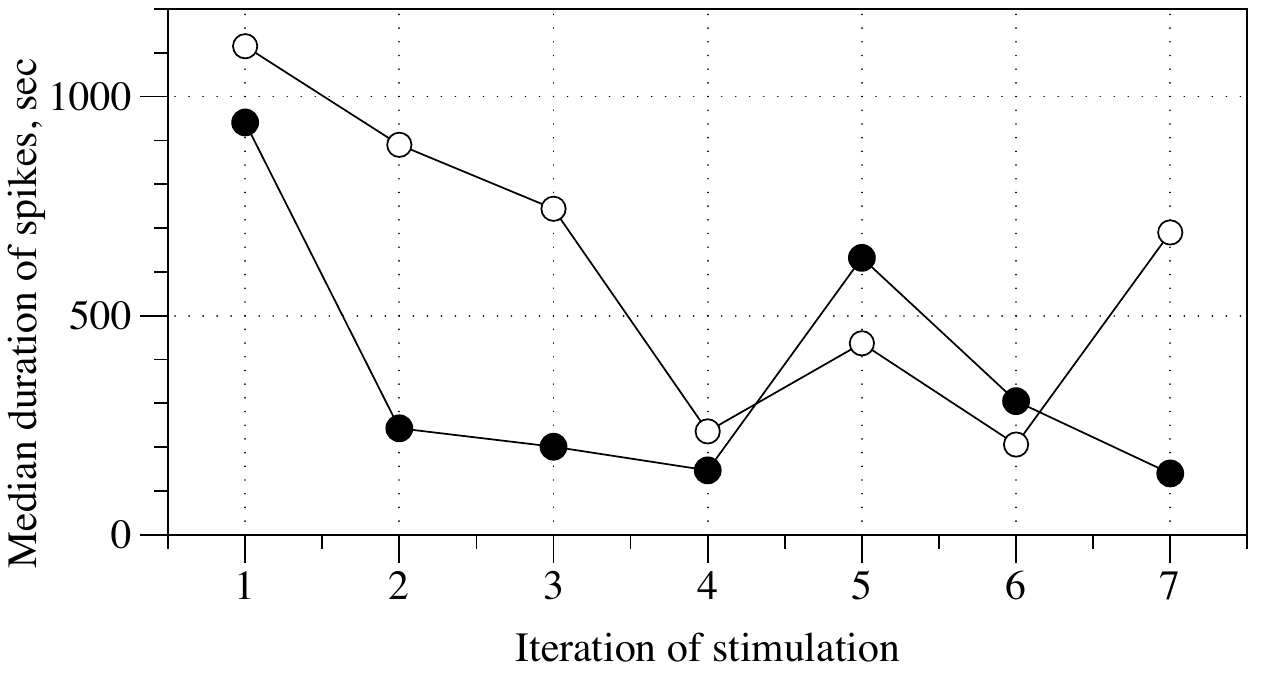}}
    \caption{Responses of living fungal blocks to stimulation as functions of iterations of stimulation. Median amplitude~(a) and median duration~(b) of ON (circles) and OFF (discs) spikes.}
    \label{fig:responsesVSiterations}
\end{figure}

Would living fungal materials habituate to the stimulation with weights? Yes, as evidenced in Fig.~\ref{fig:responsesVSiterations}. Amplitudes of ON and OFF spikes decline with iterations of stimulation as shown in Fig.~\ref{fig:responsesVSiterations}a. The duration of spikes also decreases, in overall, with iterations of stimulation,  Fig.~\ref{fig:responsesVSiterations}b, albeit not monotonously.  

\begin{figure}[!tbp]
    \centering
    \includegraphics[width=0.99\textwidth]{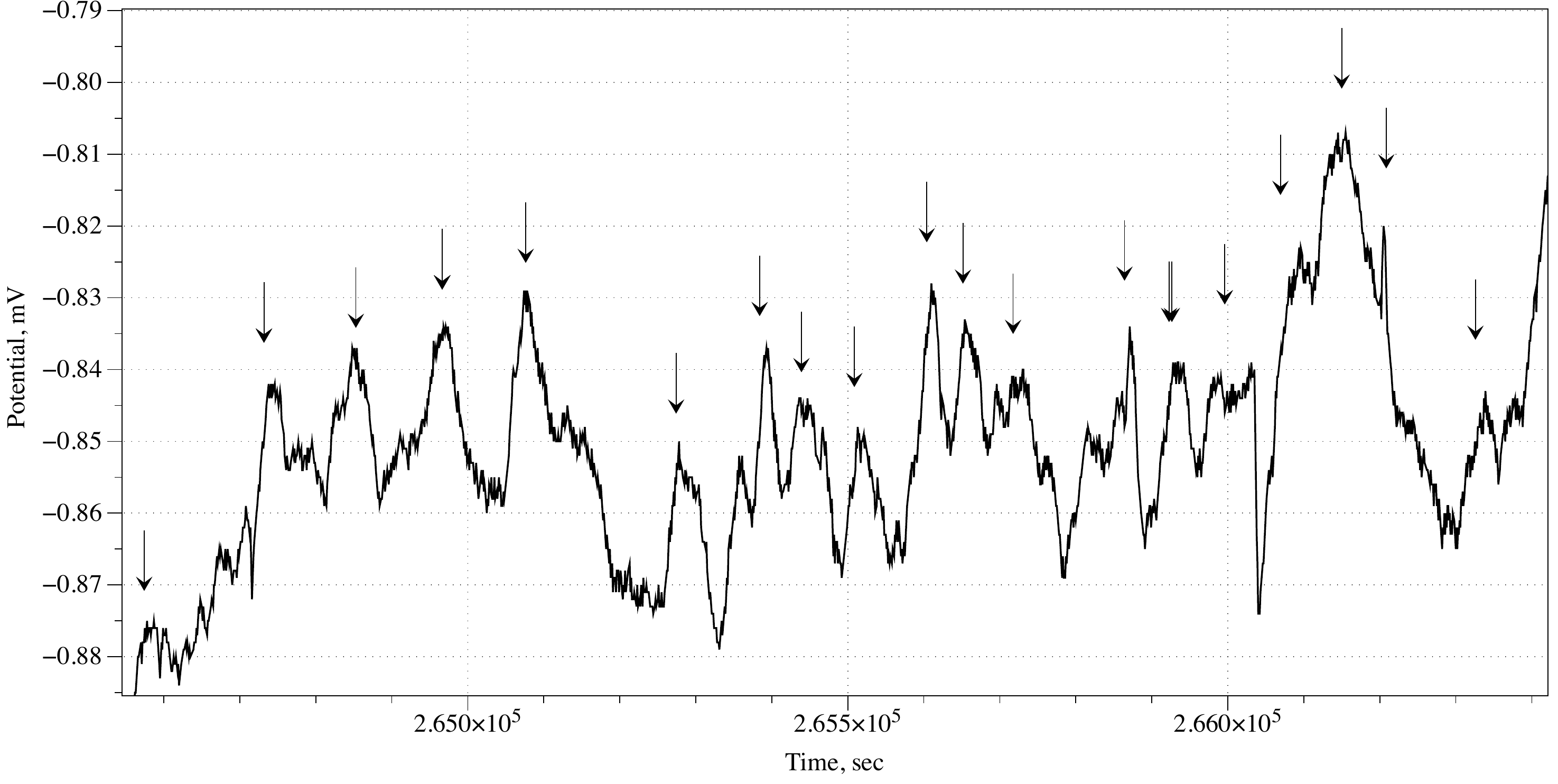}
    \caption{Example of spiking activity of living mycelium composite. The spikes of electrical potential are shown by arrows.}
    \label{fig:exampleSpiking}
\end{figure}

As previously demonstrated in~\cite{olsson1995action,adamatzky2018spiking,adamatzky2018towards} mycelium networks exhibit action-potential like spikes. An example of spiking activity recorded in present experiments is shown in Fig.\ref{fig:exampleSpiking}. The spikes have an average amplitude of 0.02~mV, $\sigma=0.01$. The amplitudes of spikes depend on a distance of an excitation wave-front from the electrodes and therefore will be ignored here, and we will focus only on frequencies of spiking. We found that median frequency of spiking of the non-stimulated fungal blocks is $\frac{1}{702}$~Hz while the fungal blocks loaded with the weights spike with a median frequency $\frac{1}{958}$~Hz. Average spiking frequency of the unloaded fungal blocks is $\frac{1}{793}$~Hz and of the loaded blocks $\frac{1}{1031}$~Hz. Thus, we can speculate that fungal blocks loaded with weights spike 1.4 times more frequently than unloaded blocks.

\section{Discussion}

We applied heavy weights to large blocks of mycelium bound composites, fungal blocks, and recorded electrical activity of the fungal blocks. We found that the fungal block respond to application and removal of the weights with spikes of electrical potential. The results complement our studies on tactile stimulation of fungal skin (mycelium sheet with no substrate) \cite{adamatzky2021fungal}: the fungal skin responds to application and removal of pressure with spikes of electrical potential. 
The fungal blocks can discern whether a weight was applied or removed because the blocks react to application of the weights with higher amplitude and longer duration spikes than the spikes responding to the removal of the weights. The fungal responses to stimulation show habituation. This is in accordance with previous studies on stimulation of plants, fungi, bacteria and protists~\cite{applewhite1975learning,fukasawa2020ecological,ginsburg2021evolutionary,boussard2019memory,yokochi1926investigation}. An additional finding was that loading of the fungal blocks with weights increase frequency of electrical potential spiking. This increase in the spiking frequency might be due to physiological responses to a mild mechanical damage caused by heavy loads; the responses involve calcium waves and lead to regeneration processes and sprouting~\cite{hernandez2015damage}.

Further studies in stimulation of living mycelium bound composites with weights could focus on studying whether shapes of the weights could be recognised by mycelium networks. A possible scenario would be to map a set of basic shapes into sets of electrical responses recorded on pairs of differential electrodes inserted into sides of the fungal blocks.

\section*{Acknowledgement}

This project has received funding from the European Union's Horizon 2020 research and innovation programme FET OPEN ``Challenging current thinking'' under grant agreement No 858132. The authors would like to acknowledge the collaboration of Mogu S.r.l. providing the living materials used in the experiments.

%\bibliography{mybibfile}

\end{document}